\documentclass[superscriptaddress,prl,twocolumn,amsmath,twocolumn,amssymb]{revtex4}
\usepackage{graphicx}
\usepackage{color}
\usepackage{eurosym}
\usepackage{dcolumn}
\usepackage{bm}
\usepackage{subfigure}
\usepackage[final]{pdfpages}
\usepackage{setspace}
\usepackage{pdfpages}
\graphicspath{{./images/}}
\usepackage{lettrine}

\begin{document}

\title{Photonic Quantum Logic with Narrowband Light from Single Atoms}

\author{Annemarie Holleczek}
\affiliation{University of Oxford, Clarendon Laboratory, Parks Road, Oxford  OX1 3PU, UK}
\author{Oliver Barter}
\affiliation{University of Oxford, Clarendon Laboratory, Parks Road, Oxford  OX1 3PU, UK}
\author{Allison Rubenok}
\affiliation{Centre for Quantum Photonics, H.\,H.\,Wills Physics Laboratory
and Department of Electrical and Electronic Engineering, University
of Bristol, Bristol BS8 1UB, UK}
\author{Jerome Dilley}
\affiliation{University of Oxford, Clarendon Laboratory, Parks Road, Oxford  OX1 3PU, UK}
\author{Peter\,B.\,R.\,Nisbet-Jones}
\affiliation{University of Oxford, Clarendon Laboratory, Parks Road, Oxford  OX1 3PU, UK}
\affiliation{now at: National Physics Laboratory, Teddington TW11 OLW, UK}
\author{Gunnar Langfahl-Klabes}
\affiliation{University of Oxford, Clarendon Laboratory, Parks Road, Oxford  OX1 3PU, UK}
\author{Graham D.\,Marshall}
\affiliation{Centre for Quantum Photonics, H.\,H.\,Wills Physics Laboratory
and Department of Electrical and Electronic Engineering, University
of Bristol, Bristol BS8 1UB, UK}
\author{Chris Sparrow}
\affiliation{Centre for Quantum Photonics, H.\,H.\,Wills Physics Laboratory
and Department of Electrical and Electronic Engineering, University
of Bristol, Bristol BS8 1UB, UK}
\affiliation{Department of Physics, Imperial College London, London SW7 2AZ, UK}
\author{Jeremy L.\,O'Brien}
\affiliation{Centre for Quantum Photonics, H.\,H.\,Wills Physics Laboratory
and Department of Electrical and Electronic Engineering, University
of Bristol, Bristol BS8 1UB, UK}
\author{Konstantinos Poulios}
\affiliation{Centre for Quantum Photonics, H.\,H.\,Wills Physics Laboratory
and Department of Electrical and Electronic Engineering, University
of Bristol, Bristol BS8 1UB, UK}
\affiliation{now at: IESL--FORTH, P.O. Box 1527, GR-71110 Heraklion, Crete, Greece}
\author{Axel Kuhn}
\email{axel.kuhn@physics.ox.ac.uk}
\affiliation{University of Oxford, Clarendon Laboratory, Parks Road, Oxford  OX1 3PU, UK}
\author{Jonathan C.\,F.\,Matthews}
\email{jonathan.matthews@bristol.ac.uk}
\affiliation{Centre for Quantum Photonics, H.\,H.\,Wills Physics Laboratory
and Department of Electrical and Electronic Engineering, University
of Bristol, Bristol BS8 1UB, UK}

\date{\today}

\begin{abstract}
\noindent 
Increasing control of single photons enables new applications of photonic quantum-enhanced technology and further experimental exploration of fundamental quantum phenomena. Here, we demonstrate quantum logic using narrow linewidth photons that are produced under nearly perfect quantum control from a single $^{87}$Rb atom strongly coupled to a high-finesse cavity. We use a controlled-NOT gate integrated into a photonic chip to entangle these photons, and we observe non-classical correlations between events separated by periods exceeding the travel time across the chip by three orders of magnitude. This enables quantum technology that will use the properties of both narrowband single photon sources and integrated quantum photonics, such as networked quantum computing, narrow linewidth quantum enhanced sensing and atomic memories.
\end{abstract}

\maketitle

\noindent 
New applications of single photons will continue to emerge from increased control of both their emission and their subsequent processing with photonic components. Today, intrinsically probabilistic photon sources, such as spontaneous parametric down conversion, are widely used for proof-of-principle photonic quantum technologies. This is because of control over properties such as entanglement~\cite{kw-prl-75-4337} and spectrum~\cite{wo-natphot-7-28}, and increasingly because of the demonstrated compatibility with integrated quantum photonics~\cite{si-natphoton-8-104}. But probabilistic sources can only generate high numbers of photons with an overhead of fast switching and optical delays~\cite{mi-pra-66-053805}. Deterministic single photon emitters circumvent this overhead whilst providing valuable capabilities such as mediating entangling operations and acting as quantum memories. Here we demonstrate that it is also possible to operate integrated quantum logic with ultra-narrowband photons emitted on-demand from single $^{87}$Rb atoms.

Integrated optics is a viable approach to control photons after they have been generated, with increasingly complex, miniature, and programable quantum circuits~\cite{po-sci-320-646, sh-nphoton-6-45, si-natphoton-8-104}.
Single photon emitters are being used with photonic quantum logic with the aim of increasing capability. For instance, sequentially emitted photons from single quantum dots have been used to measure the logical truth table of an on-chip controlled-NOT gate (CNOT)~\cite{po-apl-100-211103} and entangled using a bulk-optical CNOT \cite{ga-prl-110-250501}; photons emitted from diamond colour centres have been manipulated with an on-chip interferometer~\cite{ke-prl-111-213603}. These emitters can be regarded as artificial atomic systems. In contrast to these, ultra-narrowband indistinguishable photons can be readily obtained on-demand from real single atoms in strong coupling to high-finesse cavities~\cite{ci-prl-78-3221,ku-prl-89-067901,Solomon13}. These systems emit mutually coherent photons~\cite{Legero04}, they have been used to generate photon-atom entanglement~\cite{Wilk07-Science} and distant atom-atom entanglement~\cite{Ritter12}, they can be used for quantum memories~\cite{kh-njp-10-073023} and they can be used to individually tailor the phase and coherence envelope of each emitted single photon~\cite{Nisbet2011, di-pra-85-023834}. We seek the benefits of both integrated quantum photonic circuits and atom-cavity photon sources.

Our demonstration operates integrated photonic quantum logic with ultra-narrow-band single photons, emitted on-demand from single $^{87}$Rb atoms coupled to a high-finesse optical cavity \cite{Nisbet2011,Nisbet2013}.  We encode qubits on each single photon, that occupies one of two optical waveguides to demonstrate two-qubit logic using a probabilistic CNOT gate~\cite{ra-pra-65-062324,ho-pra-66-024308} integrated within a silica-on-silicon chip~\cite{po-sci-320-646}. We verify that for successful gate operation, the two qubits become entangled. The narrowband emission means the photons have ultra-long coherence length which manifests in our measurements as non-classically correlated detection events that are up to three orders of magnitude further apart than the time needed for light to travel across the chip. This agrees with previous measurements of time-resolved Hong-Ou-Mandel interference of long photons~\cite{Legero04}, and from which we conclude that we entangle two ultra-narrowband qubits that could be used for quantum information protocols~\cite{ko-rmp-79-135}.


\begin{figure*}[t!]
\includegraphics[width = 1.9\columnwidth]{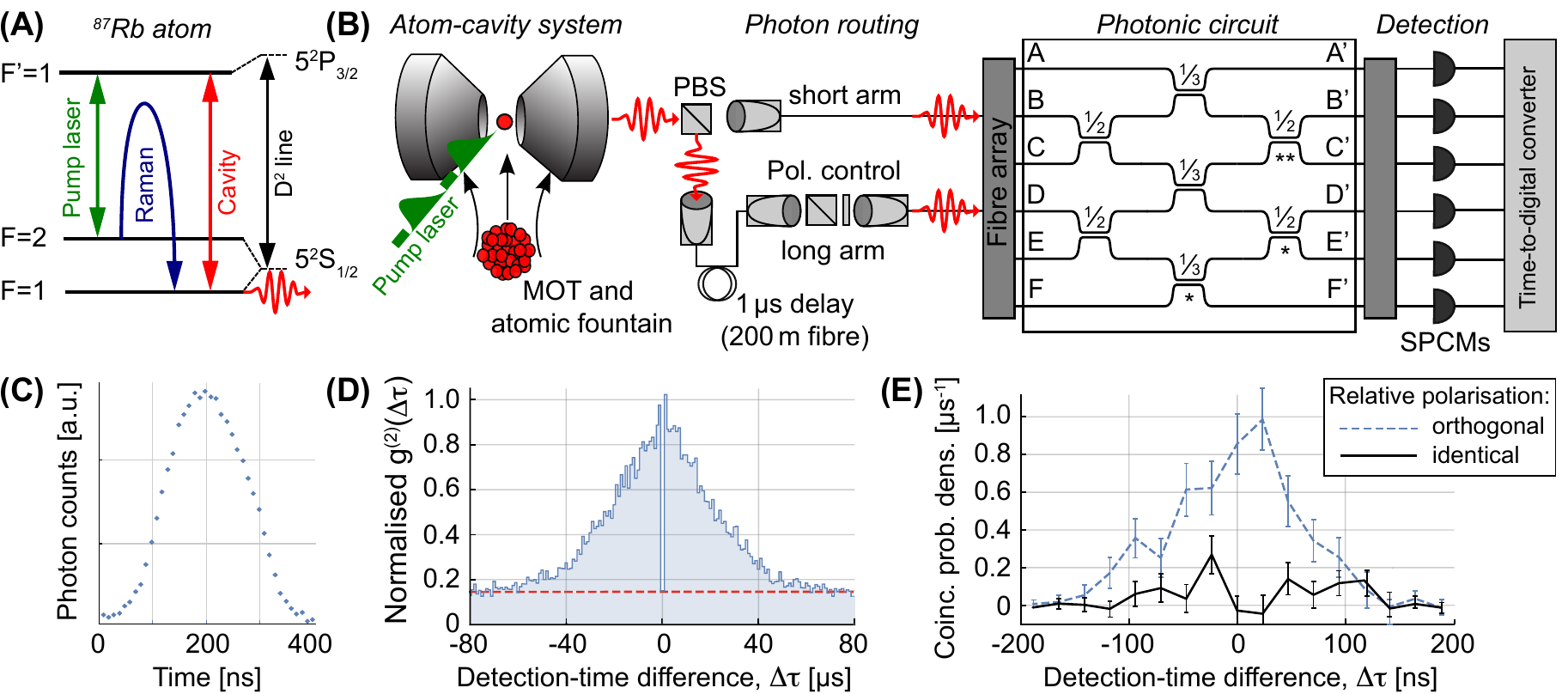}
\caption{
{Hybrid atom-cavity and photonic chip system: Vacuum-stimulated Raman transitions (A) between hyperfine ground states in $^{87}$Rb control the production of single photons.  An atomic fountain injects the atoms into a high-finesse optical cavity (B). Photons from the cavity are delayed to simultaneously enter a photonic circuit, with single-photon counting modules (SPCMs) registering the photons at all outputs with a time-to-digital converter. The spatio-temporal envelope of the photons (C) is nearly symmetric on a 400\,ns long finite support. (D) The second-order correlation $g^{(2)}(\Delta\tau)$, with temporal envelope restricted due to the finite atom-cavity $60\,\mu$s interaction time. (E) The time-resolved HOM interference pattern of two photons arriving at directional coupler ``$\ast\ast$", showing a visibility of $85(\pm 5)\%$.
}
\label{fig:cavitysetup}}
\end{figure*}

The experimental arrangement is illustrated in Fig.\,\ref{fig:cavitysetup}. 
Single photons of 60\,m (200\,ns) coherence length are derived from a strongly coupled atom-cavity system. This is accomplished with a coherently controlled vacuum-stimulated Raman transition in $^{87}$Rb with a repetition rate of 1\,MHz and an efficiency $>60\%$ \cite{ku-prl-89-067901,Nisbet2011,Nisbet2013,Solomon13}. The source operates intermittently for periods of up to $60\,\mu$s due to the stochastic delivery of atoms to the cavity with an atomic fountain. 
To obtain pairs of photons, a polarising beam splitter (PBS) directs the unpolarised photon stream into two paths of different optical length, chosen to delay one of the photons by the $1\,\mu$s period of the photon-generation sequence. 
This happens at random, so that two consecutively emitted photons are simultaneously available as a photon pair with a likelihood of $25\%$. 
Free space polarisation optics give control over the polarisations of the photons input to the chip. 

The photonic circuit, shown in Fig.\,\ref{fig:cavitysetup}(B), is a network of single-mode waveguide directional couplers designed to operate with near-infrared (NIR) photons and fabricated lithographically using germanium and boron doped silica on a silicon substrate~\cite{po-sci-320-646}. The buried square $3.5\,\mu$m$\,\times3.5\,\mu$m waveguides of refractive index contrast $\Delta n= 0.5$ support only the fundamental mode at 780\,nm. For nm bandwidth NIR photons generated via spontaneous parametric down conversion, 
with a coherence length in the $100\,\mu$m range, 
the full quantum process for single- and two-qubit logic using exactly this architecture has been characterised~\cite{sh-nphoton-6-45}. The input and output facets of the chip are glued with an optical adhesive to polarisation maintaining fibre arrays to simplify coupling into and out of the chip. The average loss across the chip from input to output fibre is 3.3\,dB. Photons emerging at the output ports of the chip get detected by silicon avalanche photodiodes with a typical quantum efficiency of $70\%$ and a time resolution of 80\,ps. Every  event is recorded and all photon-photon coincidence statistics are extracted from this data.

\begin{figure*}[tp!]
\parbox{1.28\columnwidth}{\centering
\includegraphics[width = 1.28\columnwidth]{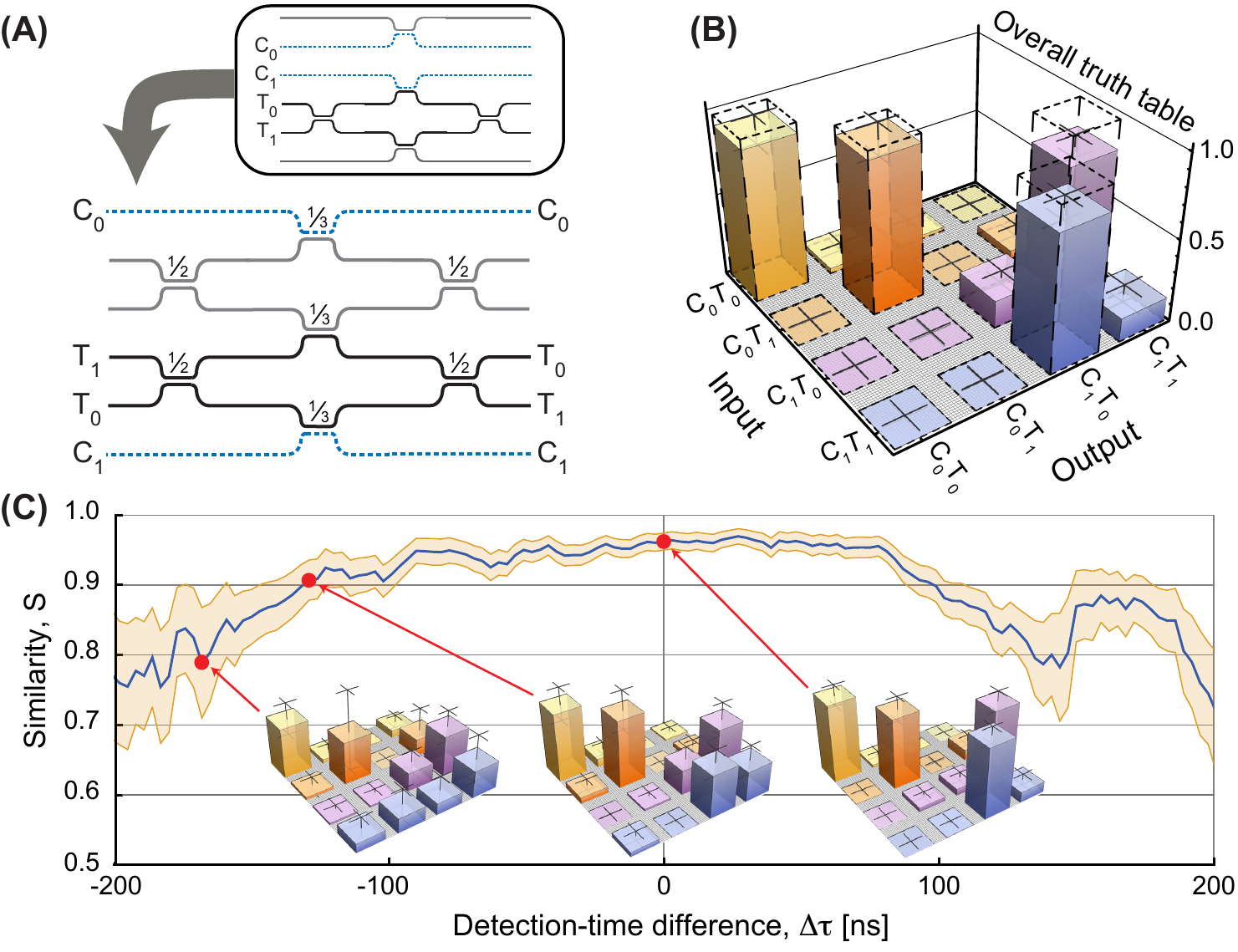}}
\hfill
\parbox{0.72\columnwidth}{\caption{
{
Linear optical CNOT gate operating on cavity photons: (A) shows the mapping of dual-rail encoded qubits to the channels of the chip we use. The overall truth table in the coincidence basis (B) is derived from 1110 pairs of control--target correlations detected up to $\pm 200$\,ns apart and measured within 20 hours. Ideal CNOT operation is presented by dotted bars. 
(C) Similarity of the truth table with the ideal CNOT as a function of the separation between detections, $\Delta\tau$. Events are considered within $\Delta\tau\pm 30\,$ns. For detections up to $100$\,ns apart, the similarity exceeds $90\%$. 
Beyond that, it drops due to noise dominating the signal. 
This is evident by correlations arising for which there is no path routing the input photons to those output channels.
}
\label{CNOTvsCNOT4}}}
\end{figure*}

The photonic chip is used in a Hanbury-Brown-Twiss configuration \cite{HanburyBrown1956} at directional couplers ``$\ast$" to characterise the photon-emission statistics of the atom-cavity system. Photons are sent along a single path into input $F$ and from the pair correlations between outputs $D'$, $E'$, and $F'$ we measure the second-order correlation function $g^{(2)}(\Delta\tau)$ shown in Fig.\,\ref{fig:cavitysetup}(D). The source operates intermittently \cite{Hennrich2004}, so that the maximum of this trace corresponds to the uncorrelated case with $g^{(2)}_{max}=1$.
 Here the finite atom-cavity interaction leads to the signal tailing off to both ends. At time delay $\Delta\tau=0$, $g^{(2)}(0)=0.15$ indicates the reduction in probability of detecting two events during a single trigger pulse. These residual correlations can be fully attributed to detector dark counts; the shot noise of which which impose an upper limit of $g^{(2)}(0) < 0.02$ to the photon stream at the one-sigma confidence level.

The mutual coherence and indistinguishability of photons is verified by time-resolved Hong-Ou-Mandel (HOM) interference of two photons \cite{ho-prl-59-2044,Legero04,Legero2006}. Photons from the long and short arms are directed into ports $A$ and $D$ respectively. Their interference at the directional coupler ``$\ast\ast$"  determines the photon-photon coincidences between detectors monitoring chip outputs $B'$ and $C'$, shown in Fig.\,\ref{fig:cavitysetup}(E), as a function of the detection-time difference.  Upon transitioning from non-interfering photons of orthogonal polarisation to interfering photons of identical polarisation, the likelihood for coincidences drops by $85(\pm 5)\%$. This large visibility of the HOM effect quantifies the degree to which our hybrid setup prepares and preserves indistinguishability of all properties of the photon pairs across their coherence time---including for overlap in the chip.

We use the photonic chip as a linear optical CNOT gate~\cite{po-sci-320-646} as shown in Fig.\,\ref{CNOTvsCNOT4}(A) \footnote{In contrast to the standard schematic~\cite{ra-pra-65-062324,ho-pra-66-024308}, here the vacuum modes get mixed on the chip. This does not affect the overall operation of the circuit as a CNOT gate.}. This gate flips the state of a target qubit conditional on the state of a control qubit. The qubits are realised in the photon pairs emitted from the cavity, with one photon guided into $C_0$ or $C_1$ and the other into $T_0$ or $T_1$. The gate's mechanism is based on two core principles of linear optical quantum circuits~\cite{ko-rmp-79-135}: single photon interference in the interferometer acting on $T_0$ and $T_1$---requiring complete circuit stability---and HOM interference, at a nominally $\eta = 1/3$ reflectivity directional coupler~\cite{ra-pra-65-062324,ho-pra-66-024308}. Operation of the gate is post-selected upon detection of one photon in $C_0$ or $C_1$ and one photon in $T_0$ or $T_1$, which occurs with probability $1/9$. The logical truth table shown in Fig.\,\ref{CNOTvsCNOT4}(B) is derived from measuring the ensemble of control-target correlated detections across the coherence envelope of the two photons. The data shown is corrected for background counts and normalised using maximum likelihood estimation (MLE) \cite{MLE}. Our results show a statistical similarity  of $S=94(\pm 1)\%$ with the ideal CNOT truth table, which increases to $S=97(\pm 1)\%$ if we account for non-ideal directional coupler reflectivities and phase shifts. We use $S=\large\sum\sqrt{p_{i} q_{i}}/\sqrt{\sum p_{i} \sum q_{i}}$, where $p_{i}$ and $q_{i}$ are elements of the measured and expected truth tables.
Due to the long coherence time of the photons, we also observe correlated detection events that are notably separated in time. From these, we determine the similarity with the expected truth table as a function of the detection-time difference, see Fig.\,\ref{CNOTvsCNOT4}(C). With the coherence length of the photons surpassing the gate dimensions (10\,mm, or 33\,ps) by three orders of magnitude, the gate operates as expected for detections up to $100$\,ns apart. Beyond that, the event rate is too small compared to noise which then dominates the truth table.

\begin{figure*}[tp!]
\parbox{1.25\columnwidth}{\includegraphics[width = 1.25\columnwidth]{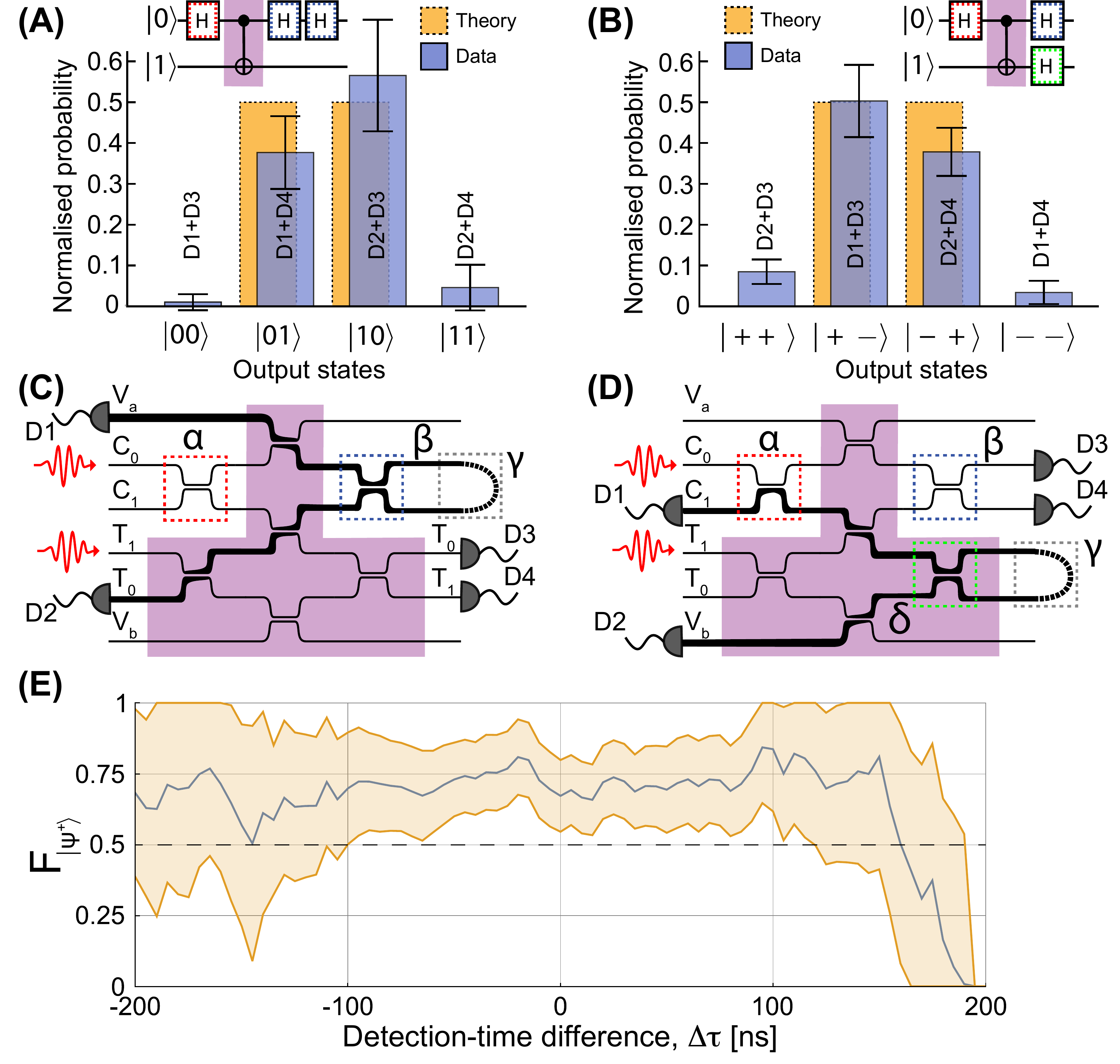}}
\hfill
\parbox{0.75\columnwidth}{\caption{
{
Non-classical qubit correlations. The circuits in (A, B) illustrate the Hadamard and CNOT gates used to measure expectation values of $Z\otimes Z$ and $X\otimes X$. The top (bottom) rail corresponds to the control (target) qubit. (A,B) The measured and ideal measurement outcomes for these two configurations. (C,D) Experimental implementations of configurations (A,D). The directional coupler $\alpha$ rotates the control qubit into the X-basis, which the CNOT (shaded region) entangles with the target qubit into the state $|\psi^+\rangle$. (C) The fibre Sagnac loop ($\gamma$) redirects the control back through through the chip along the bold path (through $\beta$ again), to rotate the control qubit back into the Z basis. (D) The Sagnac loop now leaves the control-qubit in the X basis and rotates the target qubit that now follows the bold path (through $\delta$ again) into the X-basis. Detections at detectors 1 or 2 (3 or 4) correspond to measurements of the $|0\rangle$ or $|1\rangle$ state for the control (target) qubit in (C) and $|+\rangle$ or $|-\rangle$ state in (D). (E) Variation of fidelity bound for $|\psi^+\rangle$ when considering only a subset of the detections, separated by $\Delta\tau \pm 50$ns.
}
\label{NonClassical}}}
\end{figure*}

A defining feature of two-qubit logic is the ability to generate entanglement from separable input states.
The combination of the first Hadamard and the CNOT in Fig.\,\ref{NonClassical} generates the maximally entangled  ${|\psi^+\rangle = \frac{1}{\sqrt{2}}(|01\rangle_{CT} + |10\rangle_{CT})}$ Bell state. We verify the presence of entanglement  by measuring correlations between Pauli operators along both the z- and x-axes  \cite{au-njp-8-226, wu-arxiv:0902.2093v2}. A Sagnac loop is connected to reuse part of the chip backwards as shown in Fig.\,\ref{NonClassical}(C) and (D), to allow measurement of the expectation values of the observables $Z\otimes Z$ and $X\otimes X$. We reconstruct the probability distributions for these two values using MLE, with the data normalised to the logical two-bit basis, shown in Fig.\,\ref{NonClassical}(A) and (B) respectively. The correlations show similarities of $97(\pm 3)\%$ and $94(\pm 2)\%$ with the ideal distributions. Using these measurements we can lower bound the quantum state fidelity to the $|\psi^+\rangle$ state by ${F_{|\psi^+\rangle} \geq \frac{1}{\sqrt{2}}(-\langle Z\otimes Z\rangle -\langle X\otimes X\rangle)}$, where any state with ${F_{|\psi^+\rangle} > 0.5}$ is entangled. Our data yields ${F_{|\psi^+\rangle} \geq  0.82(\pm 0.10)}$ \footnote{The non-ideal directional coupler reflectivities and phases in the interferometers preclude us from measuring $F_{|\psi^+\rangle}  \geq 0.87$.}. Much like the similarity shown in Fig.\,2(C), the degree of entanglement is largely insensitive to the detection-time difference.
We are able to observe non-classical correlations between pairs of photons that are projected onto states that could not have occupied the optical chip simultaneously. This shows that quantum interference is unaffected by photon localisation in time or space that one could otherwise associate with the two separate photon detections.

With our present work we have shown the reliable operation of two-qubit linear optical quantum gates and  generated photon-photon entanglement  \cite{ko-rmp-79-135} applied to photons emitted from a single atom strongly coupled to a cavity. 
Immediately, this new platform can be used for few-photon experiments that exploit simultaneously the capabilities of integrated quantum photonics and atom-cavity systems. Improving overall system efficiency and deterministic loading of atoms into cavities will increase the capability of this system to larger photon-number. The ultra-long coherence length of these photons exceeds the very short dimensions of the linear optical setup by several orders of magnitude. Nonetheless neither the photonic chip's ability to operate as quantum logic circuit nor the measurement of non-classical correlations due to entanglement are significantly affected. This  provides a new avenue to quantum technology that utilises narrow linewidth photons---such as quantum metrology of atomic systems~\cite{wo-natphot-7-28}---and it opens the way to large-scale networked arrays of atom-cavity systems to deterministically generate photons and act as quantum memories~\cite{ci-prl-78-3221,kh-njp-10-073023}, combined with linear optics to exchange and process quantum information. Our hybrid approach promises to be highly flexible, enabling the study of ultrafast photonics~\cite{bo-prl-108-053601} using photons of long and accessible coherence length and observing the effects of time-varying internal phases on the fidelity and errors of photonic quantum circuits \cite{Legero04,Nisbet2013}.

\footnotesize{
{\bf\em Acknowledgements:} 
We thank A. Politi for his efforts on the design of the photonic chip.
This work was supported by EPSRC, ERC, BBOI, PHORBITECH, QUANTIP, US Army Research Office (ARO) Grant No. W911NF-14-1-0133, U.S. Air Force Office of Scientific Research (AFOSR) and the Centre for Nanoscience and Quantum Information (NSQI). 
G.D.M. acknowledges the FP7 Marie Curie International Incoming Fellowship scheme. 
J.L.O'B. acknowledges a Royal Society Wolfson Merit Award and a Royal Academy of Engineering Chair in Emerging Technologies. 
A.K. acknowledges EPSRC support through the quantum technologies programme (NQIT hub) and by the German Research Foundation (DFG, RU 635). 
J.C.F.M. was supported by a Leverhulme Trust Early Career Fellowship. 
The authors are grateful to $\acute{\epsilon}\nu\alpha\gamma\rho o\nu$  $\rho\alpha\kappa\acute{\iota}$ for stimulating discussion and to D. Stuart and T. Barrett for extensive proof-reading and their most helpful comments.
}

\end{document}